\documentclass[namedreferences,hyperref,optionalrh]{spr-sola}
\usepackage{graphicx} 
\usepackage{latexsym}
\usepackage{amscd}
\usepackage{amsmath}
\usepackage{amssymb}
\usepackage{verbatim}
\usepackage{color}
\usepackage{indentfirst}
\usepackage{import}
\usepackage[utf8]{inputenc}
\usepackage{pdfpages}
\usepackage{transparent}
\usepackage{xifthen}

\newcommand{\incfig}[1]{%
    \def\svgwidth{\columnwidth}
\begingroup%
  \makeatletter%
  \providecommand\color[2][]{%
    \errmessage{(Inkscape) Color is used for the text in Inkscape, but the package 'color.sty' is not loaded}%
    \renewcommand\color[2][]{}%
  }%
  \providecommand\transparent[1]{%
    \errmessage{(Inkscape) Transparency is used (non-zero) for the text in Inkscape, but the package 'transparent.sty' is not loaded}%
    \renewcommand\transparent[1]{}%
  }%
  \providecommand\rotatebox[2]{#2}%
  \newcommand*\fsize{\dimexpr\f@size pt\relax}%
  \newcommand*\lineheight[1]{\fontsize{\fsize}{#1\fsize}\selectfont}%
  \ifx\svgwidth\undefined%
    \setlength{\unitlength}{623.62204724bp}%
    \ifx\svgscale\undefined%
      \relax%
    \else%
      \setlength{\unitlength}{\unitlength * \real{\svgscale}}%
    \fi%
  \else%
    \setlength{\unitlength}{\svgwidth}%
  \fi%
  \global\let\svgwidth\undefined%
  \global\let\svgscale\undefined%
  \makeatother%
  \begin{picture}(1,1.13636364)%
    \lineheight{1}%
    \setlength\tabcolsep{0pt}%
    \put(0,0){\includegraphics[width=\unitlength]{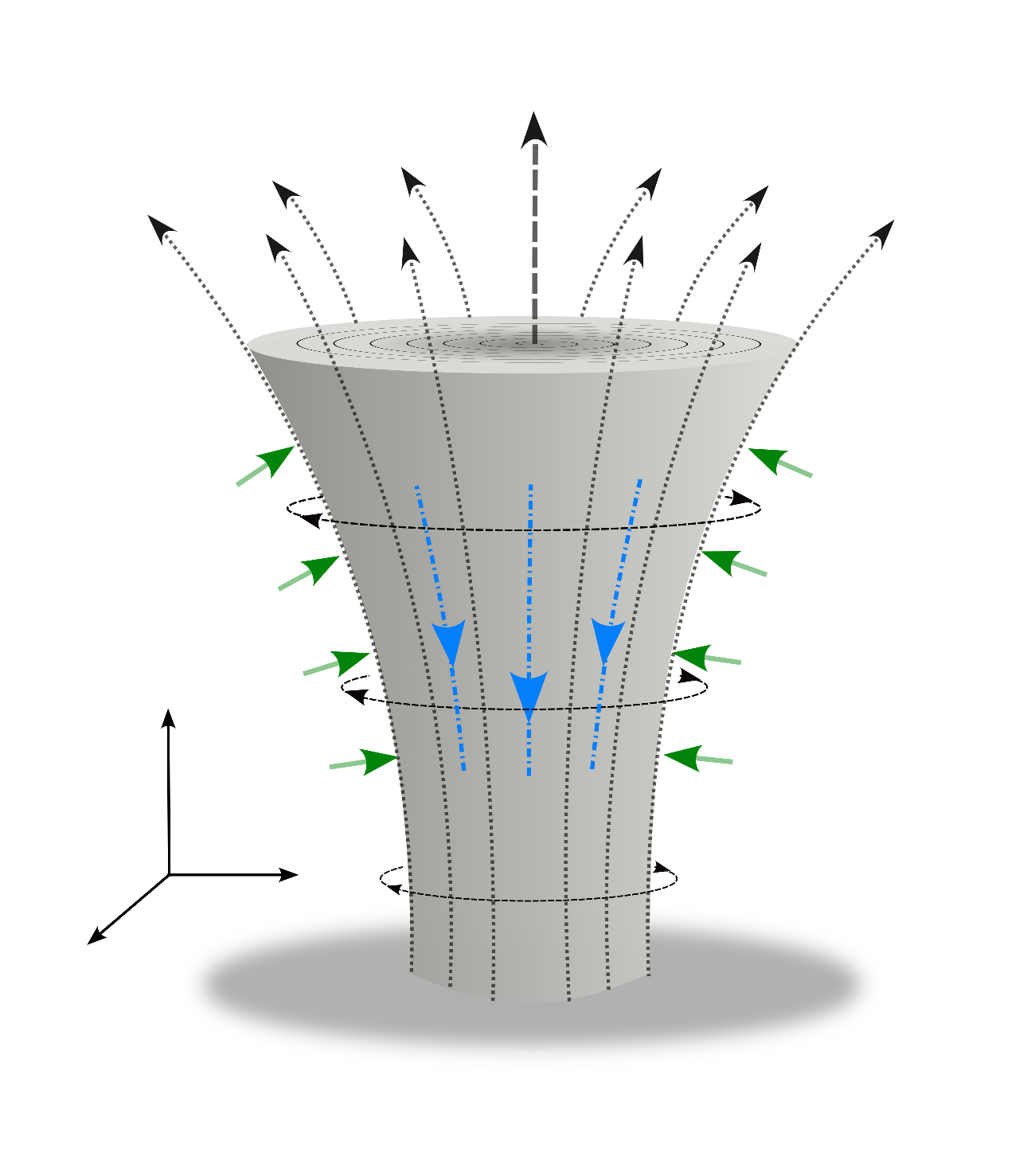}}%
    \put(0.49540951,0.35127317){\color[rgb]{0.00784314,0.00784314,0.00784314}\makebox(0,0)[lt]{\lineheight{1.25}\smash{\begin{tabular}[t]{l}\textbf{$\bold{E_{||}}$}\end{tabular}}}}%
    \put(0.08999881,0.18480315){\color[rgb]{0.00784314,0.00784314,0.00784314}\makebox(0,0)[lt]{\lineheight{1.25}\smash{\begin{tabular}[t]{l}\textbf{X}\end{tabular}}}}%
    \put(0.2902209,0.30311619){\color[rgb]{0.00784314,0.00784314,0.00784314}\makebox(0,0)[lt]{\lineheight{1.25}\smash{\begin{tabular}[t]{l}\textbf{Y}\end{tabular}}}}%
    \put(0.16078439,0.2505326){\color[rgb]{0.00784314,0.00784314,0.00784314}\makebox(0,0)[lt]{\lineheight{1.25}\smash{\begin{tabular}[t]{l}\textbf{0}\end{tabular}}}}%
    \put(0.53010486,1.01647642){\color[rgb]{0.00784314,0.00784314,0.00784314}\makebox(0,0)[lt]{\lineheight{1.25}\smash{\begin{tabular}[t]{l}\textbf{$\bold{B_{z}}$}\end{tabular}}}}%
    \put(0.7398321,0.62438115){\color[rgb]{0.00784314,0.00784314,0.00784314}\makebox(0,0)[lt]{\lineheight{1.25}\smash{\begin{tabular}[t]{l}\textbf{$\bold{B_{\phi}}$}\end{tabular}}}}%
    \put(0.70489471,0.43871139){\color[rgb]{0.00784314,0.00784314,0.00784314}\makebox(0,0)[lt]{\lineheight{1.25}\smash{\begin{tabular}[t]{l}\textbf{$\bold{Chromosphere}$}\end{tabular}}}}%
    \put(0.74592766,0.96560725){\color[rgb]{0.00784314,0.00784314,0.00784314}\makebox(0,0)[lt]{\lineheight{1.25}\smash{\begin{tabular}[t]{l}\textbf{$\bold{Corona}$}\end{tabular}}}}%
    \put(0.70355113,0.07444892){\color[rgb]{0.00784314,0.00784314,0.00784314}\makebox(0,0)[lt]{\lineheight{1.25}\smash{\begin{tabular}[t]{l}\textbf{$\bold{Photosphere}$}\end{tabular}}}}%
    \put(0.23691356,0.52756637){\color[rgb]{0.00784314,0.00784314,0.00784314}\makebox(0,0)[lt]{\lineheight{1.25}\smash{\begin{tabular}[t]{l}\textbf{$\bold{V_{n}}$}\end{tabular}}}}%
    \put(0.17563674,0.45838796){\color[rgb]{0.00784314,0.00784314,0.00784314}\makebox(0,0)[lt]{\lineheight{1.25}\smash{\begin{tabular}[t]{l}\textbf{Z}\end{tabular}}}}%
  \end{picture}%
\endgroup%

}
\usepackage{color}           

\renewcommand{\vec}[1]{{\mathbfit #1}}

\chardef\us=`\_

\begin{document}

\begin{frontmatter}
\title{Alfvén Pulse in a Chromospheric Magnetic Tube and Generation of the Super-Dreicer Electric Field}

\author[addressref={aff1,aff2},email={n.emelyanov@ipfran.ru}]{\inits{N.A.}\fnm{N.A.}~\snm{Emelyanov}\orcid{0009-0009-3931-9602}}
\author[addressref=aff1,email={kochar@ipfran.ru}]{\inits{Vl.V.}\fnm{Vl.V.}~\snm{Kocharovsky}\orcid{0000-0002-0487-4931}}

\address[id=aff1]{A.V. Gaponov-Grekhov Institute of Applied Physics of the Russian Academy of Sciences, Ulianova str. 46, Nizhny Novgorod, 603950 Russia}
\address[id=aff2]{Pulkovo Astronomical Observatory, Russian Academy of Sciences, Pulkovo chaussee 65, Saint Petersburg, 196140 Russia}

\runningauthor{N.A. Emelyanov, Vl.V. Kocharovsky}
\runningtitle{{Alfvén Pulse in a Chromospheric Magnetic Tube}}

\begin{abstract}
A self-similar solution of the linearised magnetohydrodynamic equations describing the propagation of the Alfvén pulse in an axially symmetric magnetic tube of variable diameter is obtained. The electric field component induced by the non-linear Alfvén wave and directed along the tube surface, i.e., accelerating particles along the magnetic field, is determined on the basis of the perturbation theory and specified to the case of a magnetic flux tube homogeneous over its cross section. For the chromospheric tubes, whose configuration is given by the barometric law of plasma pressure decrease, the conditions for achieving the super-Dreicer electric field limit necessary to drive the accelerated high-energy electrons into the coronal part of the loop are established. 

\end{abstract}
\keywords{Waves, Alfven, Propagation; Energetic Particles, Acceleration; Dreicer field; Flares; Chromosphere}
\end{frontmatter}
\section{Introduction}
     \label{sec1} 
\par Acceleration of charged particles in the chromospheric and coronal plasma is one of the main channels for free magnetic energy release during solar flares (\citealp{millerCriticalIssuesUnderstanding1997,aschwandenParticleAccelerationKinematics2002, benzFlareObservations2017}). Measurements of radiation intensity in the ultraviolet and X-ray bands during the most powerful flares indicate that a large number of particles are involved in the acceleration process, comparable to or even exceeding the total number of particles contained in a volume on the order of the volume of the magnetic loop coronal part (\citealp{hoyngHighTimeResolution1976,melroseTransferEnergyPotential2013, melroseBulkEnergizationElectrons2014,fleishmanSolarFlareAccelerates2022}) (the so-called "number problem"). This is difficult to explain within the framework of standard models, where the acceleration process takes place in a small region of the coronal part of the flare loop (see, e.g., \citealp{Priest2002TheMN,Shibata2011SolarFM,Toriumi2019FlareproductiveAR,gritsykModernAnalyticModels2023}).
\par The injection of particles from the chromospheric footpoints into the upper part of the loops potentially solves the above problem and explains the available data. This scenario is confirmed in particular by observations showing that particle acceleration and plasma heating can take place directly in the chromosphere, where the plasma density is much higher (see, e.g., \citealp{jiOBSERVATIONULTRAFINECHANNELS2012, sharykinFINESTRUCTUREFLARE2014}). However, it requires the presence of a strong electric field that exceeds the so-called Dreicer limit (\citealp{dreicerElectronIonRunaway1959,dreicerElectronIonRunaway1960}).
\par A scheme, in which strong electric field can be generated by a non-linear Alfvén pulse resulting from the development of the magnetic Rayleigh-Taylor instability at the chromospheric footpoint of the magnetic loop, was proposed by \cite{zaitsevParticleAccelerationPlasma2015a,zaitsevRayleighTaylorInstability2016a}. However, these works did not take into account the change in the cross section of the magnetic tube along which the Alfvén wave propagates.
\par In the present work we consider a more general formulation of the problem, not limited to the cylindrical shape of the magnetic tube and some other approximations made in the work of \cite{zaitsevParticleAccelerationPlasma2015a}, and show that a change in the geometry of the flux tube is essential for the dynamics of the Alfvén pulse and the generation of the super-Dreicer electric field. To this end, the linear and non-linear features of wave propagation are studied analytically, and the conditions for the transition to super-Dreicer acceleration regime are found in the model of barometric  law of plasma pressure decrease.
\par The paper is structured as follows. In Section \ref{sec2}, the propagation of the Alfvén pulse in an axially symmetric magnetic tube with arbitrary plasma parameters is considered in the framework of single-fluid ideal magnetic hydrodynamics (MHD) in a linear approximation, and a self-similar solution is found for the case of a longitudinal component of the magnetic field homogeneous over the tube cross-section. Section \ref{sec3} is devoted to the non-linear features of the obtained solution and to its representation in the case of a barometric model of the chromospheric part of the magnetic loops. Section \ref{sec4} contains an estimation of the non-linear electric field component responsible for the acceleration of particles along the magnetic field lines of the initial flux tube, based on the self-similar solution found for the Alfvén pulse. Section \ref{sec5} is a Conclusion.   
\section{Alfvén Pulse Propagation in an Axially Symmetric Magnetic Tube}
    \label{sec2}
\par In the article of  \cite{zaitsevParticleAccelerationPlasma2015a}, the propagation of an Alfvén wave in a circular cylindrical magnetic tube of constant diameter was studied and reduced to the description of a one-dimensional string oscillations. As will be shown below, a similar reduction to a simple problem is also possible in the case of a more general geometry with a variable magnetic tube cross-section and arbitrary plasma parameters. Earlier, the problem of this kind with use of eigenmodes theory was discussed in works of \cite{browningStructureUntwistedMagnetic1982,rudermanTransverseOscillationsLongitudinally2008,verthMAGNETOSEISMOLOGYEIGENMODESTORSIONAL2010,solerPropagationTorsionalAlfven2017}.
\par The configuration under consideration occurs naturally near regions of enhanced magnetic field strength (magnetic plugs), where the vertical component of the field varies with height. It may be an element of larger-scale magnetic structures in the solar atmosphere (the lower parts of magnetic loops, spicules, coronal holes, etc.)  (\citealp{hollwegAlfvenWavesSolar1978, hollwegAlfvenWavesSolar1981, murabitoUnveilingMagneticNature2020}) and an independent formation as well (\citealp{wedemeyer-bohmMagneticTornadoesEnergy2012,tziotziouVortexMotionsSolar2023}). Analytical calculations and observations show that magnetic tubes are the channel for the transport of MHD waves, including Alfvén waves, (see, e.g., \citealp{hollwegAlfvenWavesSolar1982,jessAlfvenWavesLower2009, fedunMHDWavesGenerated2011, depontieuUBIQUITOUSTORSIONALMOTIONS2012, srivastavaHighfrequencyTorsionalAlfven2017,liuEvidenceUbiquitousAlfven2019,battagliaAlfvenicNatureChromospheric2021}) from the upper layers of the photosphere to the middle chromosphere and transition region. There, MHD waves, damped in the non-linear regime, can transfer their energy to the plasma and thus act as one of the possible sources of solar corona heating (\citealp{hollwegAlfvenWavesSolar1981,grantAlfvenWaveDissipation2018,yadavSimulationsShowThat2020}). Therefore, a detailed analysis of MHD oscillations in the variable cross-section region of chromospheric magnetic tubes seems necessary. 
\par We consider the propagation of an Alfvén wave in an axially symmetric tube and show that, under sufficiently weak constraints, one can obtain a self-similar solution in the form of running pulses whose shape change is determined by geometrical factors of the initial configuration of the magnetic tube. In purpose to get analytical estimations (see Section \ref{sec4}) we limit our study by using ideal single-fluid magnetic hydrodynamics theory. Though the consideration of Alfvén waves dissipation and coupling with other modes could require more general approaches (\citealp{solerMagnetohydrodynamicWavesPartially2024}), it is not the subject of the current work.
\par Let us write down the general ideal-MHD equations (\citealp{priestBasicEquationsMagnetohydrodynamics2000}):
\begin{subequations}
\label{Gen_MHD_eq}
    \begin{align}
    \label{eq:MHD1.1} 
    \dfrac{\partial \vec{B}}{\partial t}=\nabla\times (\vec{v}\times \vec{B}),\\
    \label{eq:MHD1.2} 
    \dfrac{d \vec{v}}{d t}=-\frac{1}{\rho}\nabla P-\frac{1}{4\pi\rho}\Big(\vec{B}\times(\nabla\times\vec{B})\Big),\\
    \label{eq:MHD1.3}
    \frac{\partial \rho}{\partial t}+ \nabla \cdot\rho \vec{v}=0,\\
    \label{eq:MHD1.4}
    \frac{dS}{dt}=0.
    \end{align} 
\end{subequations}
Here the operator ${d}/{dt}={\partial}/{\partial t}+(\vec{v}\cdot\nabla)$ is introduced, $\vec{B}$ is the magnetic field induction, $\vec{v}$ the plasma flow velocity, $\rho$ the plasma density, $P$ the plasma pressure, $S$ the plasma entropy.  
\par We will assume the plasma to be incompressible, i.e., the condition 
 $\nabla\cdot\vec{v}=0$ is fulfilled. For an axially symmetric magnetic tube, all these quantities have no dependence on the azimuthal angle  $\phi$ and, consequently, all derivatives of this variable are equal to zero: ${\partial}/{\partial \phi}=0$.
In the linear approximation, we will assume that there is no radial plasma flow: $v_{r}\equiv 0$ (the validity of this approximation is discussed in Section \ref{sec3}). Then, from the incompressibility condition $\nabla\cdot\vec{v}=0$, it follows that the longitudinal component of the plasma flow velocity is also absent, $v_{z}\equiv0$, and from the equation for magnetic induction \ref{Gen_MHD_eq}(a) we have ${\partial B_{z}}/{\partial t} = 0 $, ${\partial B_{r}}/{\partial t} = 0$, i.e., the longitudinal and radial components of the magnetic field, $B_{z}(r,z)$ and $B_{r}(r,z)$, are completely determined by the initial values, as well as the plasma density $\rho(r,z)$.
\par This leaves only two varying azimuthal components, $v_{\phi}$ and $B_{\phi}$, which determine the dynamics of the Alfvén wave.
The MHD equations for these components are the following:  
\begin{align}
\begin{cases}
    \label{eq:MHD2.1} 
    \dfrac{\partial B_{\phi}}{\partial t}=B_{z}\dfrac{\partial v_{\phi}}{\partial z}+B_{r}\dfrac{\partial v_{\phi}}{\partial r}-v_{\phi}\dfrac{B_{r}}{r},\\
    \dfrac{\partial v_{\phi}}{\partial t}=\dfrac{1}{4\pi\rho}\Big(B_{z}\dfrac{\partial B_{\phi}}{\partial z}+B_{r}\dfrac{\partial B_{\phi}}{\partial r}+B_{\phi}\dfrac{B_{r}}{r}\Big).    
\end{cases}
\end{align}
Let us differentiate the first equation by time and substitute the second equation into it. The result is 
\begin{align}
    \dfrac{\partial^{2} B_{\phi}}{\partial t^{2}}=\frac{B_{z}^{2}}{4\pi\rho}\dfrac{\partial^{2} B_{\phi}}{\partial z^{2}}+2\frac{B_{z}B_{r}}{4\pi\rho}\dfrac{\partial^{2} B_{\phi}}{\partial z\partial r}+\frac{B_{r}^{2}}{4\pi\rho}\dfrac{\partial^{2} B_{\phi}}{\partial r^{2}} \nonumber\\
    +\frac{B_{z}}{4\pi}\dfrac{\partial}{\partial z}\Big(\frac{B_{z}}{\rho}\Big)\dfrac{\partial B_{\phi}}{\partial z}+\frac{B_{r}}{4\pi}\dfrac{\partial}{\partial r}\Big(\frac{B_{z}}{\rho}\Big)\dfrac{\partial B_{\phi}}{\partial z}+\frac{B_{z}}{4\pi}\dfrac{\partial}{\partial z}\Big(\frac{B_{r}}{\rho}\Big)\dfrac{\partial B_{\phi}}{\partial r}+\frac{B_{r}}{4\pi}\dfrac{\partial}{\partial r}\Big(\frac{B_{r}}{\rho}\Big)\dfrac{\partial B_{\phi}}{\partial r} \nonumber\\
    +\frac{B_{z}}{4\pi}\dfrac{\partial}{\partial z}\Big(\frac{B_{r}}{r\rho}\Big)B_{\phi}+\frac{B_{r}}{4\pi}\dfrac{\partial}{\partial r}\Big(\frac{B_{r}}{r\rho}\Big)B_{\phi}-\frac{B_{r}^{2}}{4\pi\rho}\frac{B_{\phi}}{r^2}.
\end{align}
\par This equation describes the propagation of an Alfvén wave in an axially symmetric magnetic tube with arbitrary dependencies ${B_{r}(r,z)}$, ${B_{z}(r,z)}$ and ${\rho(r,z)}$. However, in its present form it is difficult to analyse. Let us choose new variables $\xi(r,z)$ and $\eta(r,z)$ so that the wave equation has the simplest form. Namely, we use the method of characteristics (\citealp{whithamLinearNonlinearWaves1974}) to write down the following characteristic equation:
\begin{align}
    B_{z}^{2}dr^2-2B_{z}B_{r}dzdr+B_{r}^{2}dz^2=0.
\end{align}
Its solution, obviously, is the equation of the line of the magnetic tube surface:
\begin{align}
   \label{field_line_eq}
   \frac{dr}{B_{r}(r,z)}=\frac{dz}{B_{z}(r,z)}.
\end{align}
With use of the divergence-free condition $\nabla \cdot\vec{B}=0$ one can introduce the magnetic vector potential $\vec{A}$ in a usual way:
\begin{align}
   \label{vec_pot}
   \vec{B}=\nabla \times \vec{A}.
\end{align}
Substituting the expression \ref{vec_pot} into the characteristic Equation \ref{field_line_eq} we get
\begin{align}
    \frac{1}{r}\dfrac{\partial (r A_{\phi})}{\partial r} dr+\dfrac{\partial A_{\phi}}{\partial z}dz=0.
\end{align}
As can be easily seen, $d(r A_{\phi})=0$ and, therefore, $ r A_{\phi}=const$,
i.e., the value $r A_{\phi}$ is the integral of the characteristic equation. Thus, there is a manifold of solutions which describe the set of nested magnetic tubes defined by its constant value $r A_{\phi}$. A particular dependence of the tube radius on the longitudinal coordinate is defined by setting the value of the integral, which selects a unique solution from the manifold: $r|_{\xi=const}=r(z)$. 
\par We can use the following new coordinates expressed by the old ones: 
\begin{subequations}
    \label{new_var}
    \begin{align}
    \xi=r A_{\phi}(r,z),\label{invar_1}\\
    \eta=z. \label{invar_2}
    \end{align}
\end{subequations}
\par After replacing the variables we have $B_{\phi}(t,r,z)=B_{\phi}(t,\xi(r,z),\eta(r,z)) \equiv u(t,\xi,\eta)$.
Note that the integral $r A_{\phi}=const$ of the characteristic Equation $\ref{field_line_eq}$ has a simple physical meaning of the magnetic flux conservation law: 
\begin{align}
    \Phi=\oint_{\gamma}\vec{A}\cdot d\vec{l}=\int_{0}^{2\pi}A_{\phi}(r,z)rd\phi=2\pi rA_{\phi}=2\pi \xi.
\end{align}
In addition, this value can be related to the $z$-component of the angular momentum of the magnetic field, $M_{z}=r ({e}/{c}) A_{\phi}$, the conservation of which is a consequence of the axial symmetry of the system.
\par In the new coordinates, the equation for the Alfvén wave can be rewritten as follows:
\begin{align}
\label{reduce_wave_eq}
    \dfrac{\partial^{2} u}{\partial t^{2}}=\frac{B_{z}^{2}}{4\pi\rho}\dfrac{\partial^{2} u}{\partial \eta^{2}}+
    \frac{B_{z}}{4\pi}\dfrac{\partial}{\partial \eta}\Big(\frac{B_{z}}{\rho}\Big)\dfrac{\partial u}{\partial \eta}+ \Big(\frac{B_{z}}{4\pi}\dfrac{\partial}{\partial \eta}\Big(\frac{B_{r}}{r\rho}\Big)-\frac{1}{r^2}\frac{B_{r}^{2}}{4\pi\rho}\Big)u.
\end{align}
We are looking for the solution of this wave equation in the following form:
\begin{align}
    u=f(t,\xi,\eta)\exp\Big({\frac{1}{2}\int^\eta \frac{1}{B_{z}} \dfrac{\partial B_{z}}{\partial \eta'} d\eta'}\Big).
\end{align}
Then, carrying out simple calculations, we obtain
\begin{align}
    \label{eq:string}
    \dfrac{\partial^{2} f}{\partial t^{2}}=\dfrac{\partial}{\partial \eta}\Big(c_{A}^{2}\dfrac{\partial f}{\partial \eta}\Big)+\frac {f}{{\tau}^{2}},
\end{align}
where $c_{A}^{2}={B_{z}^{2}}/{4\pi\rho}$ is the square of the Alfvén velocity, and the parameter ${\tau}$ has the dimension of time and is defined by the sign-indefinite expression
\begin{align}
    \frac{1}{{\tau}^2}=\dfrac{\partial}{\partial \eta}\Big(\frac{c_{A}^{2}}{L}\Big)-\frac{c_{A}^{2}}{L^2}.
\end{align}
\par Here, the value $L$  has the dimension of length and determines the characteristic scale of the longitudinal change of the function $\varkappa(\xi,\eta)={(\vec{B}\cdot\vec{S})}/{\Phi}$:
\begin{align}
\label{L_func}
    \frac{1}{L(\xi,\eta)}=\frac{1}{2\varkappa(\xi,\eta)}\dfrac{\partial \varkappa (\xi, \eta)}{\partial \eta}.
\end{align}
\par The introduced function $\varkappa$ characterises the ratio of the edge value of the longitudinal component of the magnetic field, $B_{z}(r)$, to its cross-sectional mean value for each magnetic tube with a fixed value of magnetic flux $\Phi$. Here $\vec{S}=\pi r^2  \hat{\bf{z}}$ is the vector cross-sectional area of the magnetic tube corresponding to a fixed value of magnetic flux $\Phi$. The unit vector $\hat{\bf{z}}$ is directed along the OZ-axis. Note that all functions in Equations \ref{reduce_wave_eq}-\ref{L_func}, including $c_{A}$, depend on the longitudinal coordinate, $\eta=z$, but also on the coordinate $\xi$.
\par Thus, in the general case, assuming only axial symmetry of the magnetic tube with plasma, the derivatives on the variable $\xi$, which now enters Equation \ref{eq:string} only as a parameter, are completely excluded by appropriate substitution of the variables. In other words, the linear part of the Alfvén wave propagation problem is reduced to the oscillations of a one-dimensional string whose properties are determined by a given initial distribution of the magnetic field components $B_{r}(r,z)$, $B_{z}(r,z)$ and the plasma density $\rho(r,z)$.
\par If the longitudinal component of the magnetic field $B_{z}$ does not depend on the radius $r$, but is an arbitrary function of the longitudinal coordinate $z$, then the condition $\nabla \cdot\vec{B}=0$ implies that there is the radial component, $B_{r}(r,z)=-({r}/{2}){\partial B_{z}(z)}/{\partial z}$.
\par On the other hand, for the value $\xi$ we have
\begin{align}
    \dfrac{\partial \xi}{\partial z}=r\dfrac{\partial A_{\phi}}{\partial z}=-rB_{r}=\frac{r^2}{2}\dfrac{\partial B_{z}}{\partial z}.
\end{align}
In this case, choosing the natural calibration of the vector potential $A_{\phi}(r,z)= (r /2)B_{z}(z)$, we get:
\begin{align}
     \xi(r,z)=\frac{r^2}{2} B_{z}(z).
\end{align}
Then, by definition, $\varkappa \equiv 1$ and, as it can be easily seen, ${1}/{\tau^2} \equiv 0$. As a result, the wave Equation \ref{eq:string} takes a simple form
\begin{align}
    \dfrac{\partial^{2} f}{\partial t^{2}}=
    \dfrac{\partial}{\partial \eta}\Big(c_{A}^{2}\dfrac{\partial f}{\partial \eta}\Big).
\end{align}
\par There are many constructive methods for solving this type of equations. In the simplest case of an adiabatically weak dependence of the Alfvén velocity $c_{A}^{2}(\eta)$ on the longitudinal coordinate, we come to the standard wave equation  
\begin{align}
    \label{red_w_eq}
    \dfrac{\partial^{2} f}{\partial t^{2}}=
    c_{A}^{2}(\eta)\dfrac{\partial^2 f}{\partial \eta^2}.
\end{align}
This simplification also assumes weak dependence of the plasma density $\rho(r)$ over the tube cross section. Though dispersion of the Alfvén velocities due to transverse plasma inhomogeneity can be an important feature and lead to generation of strong electric fields, which produce plasma heating and particles acceleration (see, e.g., \citealp{tsiklaurSakaiPICAlfven2005, tsiklauriMechanismElectricMHD2006, McClements_2009, bianParallelElecField2011}), we do not consider it in the present work.  
\par Finally, using the well-known d'Alambert method, we obtain that for a given initial perturbation of the field $\psi(\xi,\eta)$, the solution of Equation \ref{red_w_eq} has the following form:
\begin{align}
    f=\frac{\psi(\xi,\eta-c_{A}(\eta)t)+\psi(\xi,\eta+c_{A}(\eta)t)}{2}.
\end{align}
\par Thus, by relating the initial value of the azimuthal component of the magnetic field $B_{\phi 0}(r,z)$, determined by the current distribution in the magnetic tube, to the initial perturbation $\psi(\xi,\eta)$ in the string oscillation problem, one can obtain a self-similar solution for the Alfvén perturbation at all subsequent times as the sum of two 'running' pulses ($+$ and $-$):
\begin{align}
    \label{self-sim. solution}
    B_{\phi}(t,r,z)=\frac{1}{2}\Big\{B_{\phi 0}\Big(r\sqrt{\frac{B_{z}(z)}{B_{z}(z- c_{A}t)}},z-c_{A}t\Big)\exp{\Big(\frac{1}{2}\int_{z- c_{A}t}^{z}\frac{1}{B_{z}}\dfrac{\partial B_{z}}{\partial z'}dz'}\Big)+\nonumber\\ B_{\phi 0}\Big(r\sqrt{\frac{B_{z}(z)}{B_{z}(z+ c_{A}t)}},z+ c_{A}t\Big)\exp{\Big(\frac{1}{2}\int_{z + c_{A}t}^{z}\frac{1}{B_{z}}\dfrac{\partial B_{z}}{\partial z'}dz'}\Big)\Big\}.
\end{align}
Note the scaling of the pre-exponential factor, $B_{\phi0}(...)$,  as a function on the cross-sectional coordinate $r$, is defined by the square root of the ratio of $z$-components of the tube magnetic field at two different vertical coordinate, $z$ and $z\pm c_{A}t$.  
\section{Non-linear Properties of the Alfvén Pulse and a Barometric Model of the Chromospheric Part of the Magnetic Loop} 
      \label{sec3}      
\par Let us estimate the non-linear corrections to the obtained solution \ref{self-sim. solution} due to the presence of the plasma flow velocity components $v_{r}$ and $v_{z}$.
According to the Euler equation projected on the OZ-axis, one can obtain
\begin{align}
\label{Euler_eq}
    \dfrac{\partial v_{z}}{\partial t}\approx-\frac{1}{\rho}\dfrac{\partial P}{\partial z}-\frac{1}{4\pi\rho}\Big(B_{r}(\dfrac{\partial B_{r}}{\partial z}-\dfrac{\partial B_{z}}{\partial r})+B_{\phi}\dfrac{\partial B_{\phi}}{\partial z}\Big).
\end{align}
Assuming that the appearance of the longitudinal component of the velocity $v_{z}$ is determined only by the action of the Alfvén pulse, we approximate Equation \ref{Euler_eq} as follows: 
\begin{align}
    \dfrac{\partial v_{z}}{\partial t}\approx-\frac{B_{\phi}}{4\pi\rho}\dfrac{\partial B_{\phi}}{\partial z}.
\end{align}
As a result, using the self-similar solution \ref{self-sim. solution}, we come to the following estimation:
\begin{align}
   | v_{z}|\approx \frac{1}{c_{A}}\frac{B_{\phi}^2}{4\pi\rho} =\Big(\frac{B_{\phi}}{B_{z}}\Big)^2c_{A} .
\end{align} 
In the case of weak perturbations, when $({B_{\phi}}/{B_{z}})^2 \ll 1$, we get $|v_{z} |\ll c_{A}$. From the incompressibility condition $\nabla\cdot\vec{v}=0$ we also have: $v_{r} \sim v_{z}( {l_{r}}/{l_{z}})$, where $l_{r}, l_{z}$ are characteristic scales of the arising flows. Thus, in the above case, for a not too large radial scale $l_{r}$, the plasma motion does not lead to significant changes in the configuration of the magnetic tube during the passage time of the Alfvén pulse and, consequently, to distortions of the pulse itself, since both velocity components are small compared to the Alfvén velocity, $v_{z}, v_{r}\ll c_{A}$, confirming the validity of the linear approximation. 
\par However, the very appearance of non-linear components of the plasma velocity as a result of the passage of the Alfvén wave leads to the appearance of non-linear components of the electric field, which can inject plasma from the chromospheric part into the coronal region of the loop and accelerate particles. The result can depend on the longitudinal profile of the tube magnetic field. This issue is discussed in more detail in the next section.
\par Note that for typical values of the Alfvén velocity $ c_{A}\sim 10^{3}$ km 
 s$^{-1}$ and a not too small ratio $({B_{\phi}}/{B_{z}})^2 \sim 10^{-1}-10^{-2}$ , which can be achieved in magnetic loops with currents $\gtrsim 10^{10}$ A and characteristic diameters of flux tubes $10^{2}-10^{3}$ km (\citealp{sharykinFINESTRUCTUREFLARE2014}), the obtained estimated value of the non-linear vertical component of the plasma velocity $v_{z}$ is of the order of $\sim 10$ km 
s$^{-1}$. This value is confirmed by the observation of chromospheric plasma injection into the coronal part of loops (see, e.g., \citealp{jiOBSERVATIONULTRAFINECHANNELS2012,depontieuUBIQUITOUSTORSIONALMOTIONS2012}).
\par Consider the propagation of the Alfvén pulse in an axially symmetric magnetic tube of a chromospheric part of a magnetic loop, where plasma density and pressure can be sufficiently high.
\par From the equilibrium condition one should expect that the configuration of the chromospheric flux tube is determined by the balance of the magnetic pressure ${B^2}/{8\pi}$ inside the tube (the pressure of the plasma itself inside the tube is considered negligibly small) and the external gas-kinetic pressure $P_{e}$ outside the tube  (see, e.g., \citealp{verthMAGNETOSEISMOLOGYEIGENMODESTORSIONAL2010,murawskiNumericalModelMHD2015}). If the longitudinal component of the magnetic field is dominant ($B_{z}\gg B_{r},B_{\phi}$), this condition has the form: ${B_{z}^2}/{8\pi}\approx P_{e}$. Suppose further that this component $B_{z}(z)$ depends only on the longitudinal (vertical) coordinate, and that the external pressure in equilibrium is given by the quasi-barometric formula 
\begin{align}
   P_{e}(z)=P_{0}\exp\big({-\int_{0}^{z}\frac{dz'}{H(z')}}\big),
\end{align}
where $H(z)={k_{B}T(z)}/({m_{i}g})$ is the scale height, $k_{B}$ the Boltzmann constant, $T(z)$ the plasma temperature, $m_{i}$ the ion mass, $g$  the acceleration of free fall at the surface of the Sun\footnote{Actually, the value of $H(z)$ varies weakly within the chromosphere (\citealp{avrettModelsSolarChromosphere2008}), remaining of the order of a few hundred kilometres.}. 
Then from the equilibrium condition we have: 
\begin{align}
    \label{B_z}
   B_{z}(z)=B_{0}\exp\big({-\int_{0}^{z}\frac{dz'}{2H(z')}}\big).
\end{align}
\par In the work of \cite{zaitsevParticleAccelerationPlasma2015a}, it was shown that for such a steady-state field configuration, by virtue of the magnetic flux conservation law, there is an exponential expansion of the magnetic tube, 
\begin{align}
        \label{radius}
        a(z)=a_{0}\exp\big({\int^{z}_{0}\dfrac{dz'}{4H(z’)}}\big).
\end{align}
Here $a(z)$ is the tube radius, so that the plasma parameter, $\beta=8\pi P/B^{2}$, is larger and smaller than unity outside, $r>a$, and inside, $r<a$, the tube, respectively. It was also shown that this configuration can become unstable with respect to the growth of the Rayleigh-Taylor mode providing the generation of the Alfvén pulse.
\par For the sake of clarity, we assume that the plasma density in the chromospheric part of the loop decreases according to the same exponential law as the external pressure. In this case, the changes in density and magnetic field are consistent, so that the Alfvén velocity $c_{A}$ varies weakly with height following a slow dependence of the plasma temperature on the longitudinal coordinate. Such a consideration is reasonable at altitudes below the transition region between the chromosphere and the corona. 
\par Under these assumptions, the approximations used in the previous section are valid. Substituting the specific dependence of the longitudinal component of the magnetic field, Equation $\ref{B_z}$, into the expression \ref{self-sim. solution} for the Alfvén perturbation, and leaving only the upward pulse, we have 
\begin{align}
    \label{expon_sol}
    B_{\phi}(t,r,z)=\frac{1}{2}\exp\big({-\int^{z}_{z-c_{A}t}\frac{dz'}{4H(z')}}\big)B_{\phi 0}\Big(r\exp\big({-\int^{z}_{z-c_{A}t}\frac{dz'}{4H(z')}}\big),z-c_{A}t\Big).
\end{align}
\par
\begin{figure}    
\centerline{\hspace*{0.015\textwidth}
         \includegraphics[width=0.515\textwidth,clip=]{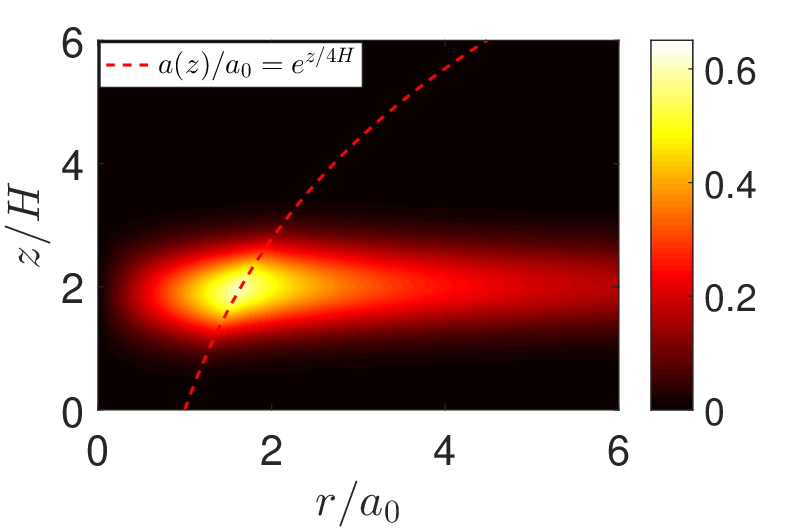}
         \hspace*{-0.03\textwidth}
         \includegraphics[width=0.515\textwidth,clip=]{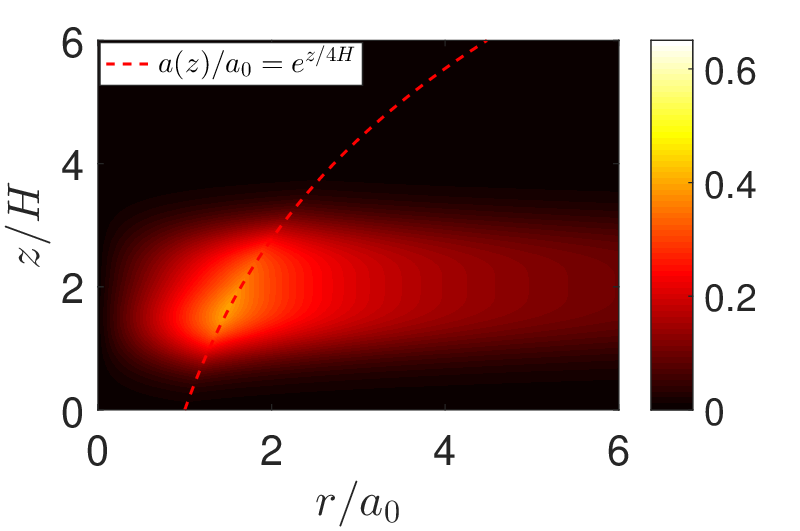}
        }
\vspace{-0.31\textwidth}   
\centerline{\Large \bf     
\hspace{0.31 \textwidth}  \color{white}{(a)}
\hspace{0.415\textwidth}  \color{white}{(b)}
   \hfill}
\vspace{0.31\textwidth}    
\centerline{\hspace*{0.015\textwidth}
         \includegraphics[width=0.515\textwidth,clip=]{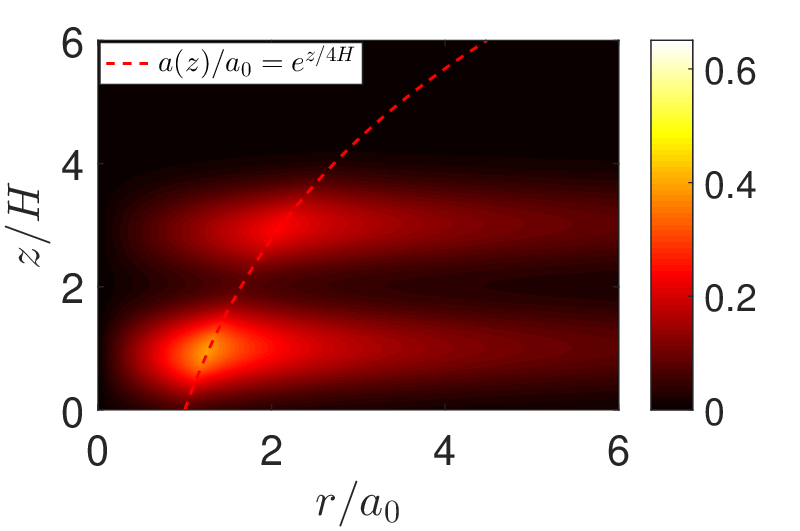}
         \hspace*{-0.03\textwidth}
         \includegraphics[width=0.515\textwidth,clip=]{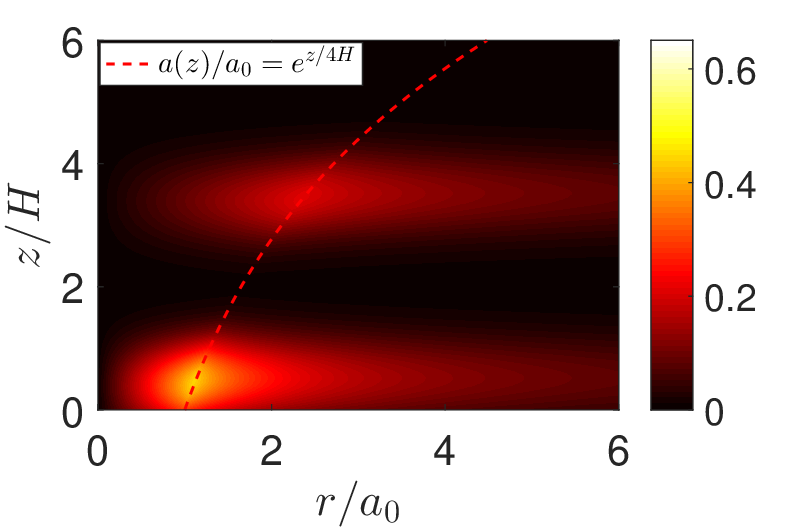}
        }
\vspace{-0.31\textwidth}   
\centerline{\Large \bf     
\hspace{0.31 \textwidth} \color{white}{(c)}
\hspace{0.415\textwidth}  \color{white}{(d)}
   \hfill}
\vspace{0.31\textwidth}    
              
\caption{Space distribution of the dimensionless magnetic field component $B_{\phi}$  of the Alfvén pulse at different times $\tilde{\tau}=0,0.5,1,1.5$ (a,b,c,d respectively), where $\tilde{\tau}={c_{A}t}/{H}$. The dashed curve is the radius of the magnetic tube.}
\label{fig:wave_propag}
\end{figure}

 Thus, the Alfvén pulse weakly changes in the layer of thickness of the order of $H$ adjacent to the photosphere as it moves toward the corona. However, it stretches significantly in the transverse direction, following the scaling of the exponential expansion of the magnetic tube, and decays exponentially as it travels a distance greater than $H$ (see Figure \ref{fig:wave_propag}).
\par According to the expression \ref{self-sim. solution}, there is also a second pulse, exponentially increasing, moving downward to the photosphere. However, the ratios ${B_{\phi}}/{B_{z}}$ and ${B_{\phi}^{2}}/({8\pi P})$ for such a pulse exponentially decrease when the distance travelled is greater than $H$, which weakens the generation of the non-linear electric field component discussed below in Section \ref{sec4}. 
\par In Figure \ref{fig:wave_propag}, the illustration of the Alfvén pulse propagation along the chromospheric tube defined by the self-similar solution \ref{self-sim. solution} is presented in the case of the exponentially expanding magnetic tube ($H=const$) with the electric current distribution homogeneous over the cross section, $j(r)=const$ for $r\leqslant a(z)$ and $j(r)=0$ for $r > a(z)$, and with a Gaussian distribution of the total current, $I(z)\sim \exp{(-{(z-z_{0})^{2}}/{\sigma^{2}}})$,  along the longitudinal $z$ - coordinate (in this figure $z_{0}=2H$, $\sigma^{2}=0. 5 H^{2}$).  
\par In the present work we have not considered possible reflection from the photosphere and transition region as well as dissipation of the Alfvén pulse. These issues and an expected power of the Alfvén pulses will be discussed elsewhere.
\section{Generation of the Accelerating Electric Field by a Non-linear Alfvén Pulse} 
      \label{sec4}      
\par The particles acceleration in chromospheric part of magnetic loops is defined to a large extent by the magnitude of the electric field induced by the Alfvén pulse. Using the ideal conductivity condition, we express the electric field strength in terms of the local plasma velocity and the magnetic induction vector:
\begin{align}
    \vec{E}=-\frac{1}{c}(\vec{v}\times\vec{B}).
\end{align}
\par
\begin{figure}[t]
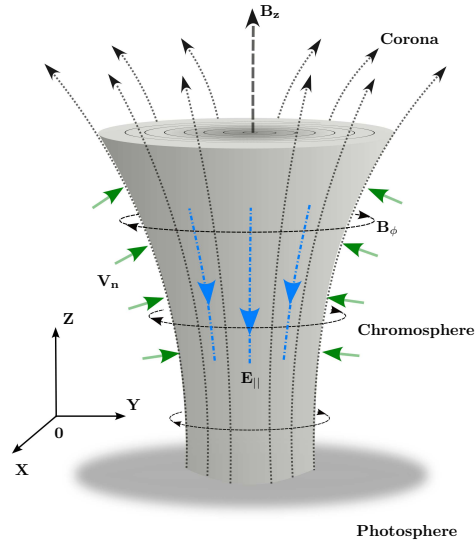
  
    \centering
    
    \scalebox{0.6}{\incfig{footpoint_tex2}}
    
    \caption{The expanding magnetic tube of the chromospheric part of the coronal loop. The black dashed arrows show $B_{\phi}$ component and the black dotted arrows depict the magnetic field lines of the flux tube surface, respectively. The green arrows indicate the normal component of the plasma flow velocity $V_{n}$, and the blue dash-dotted arrows show the non-linear electric field component $E_{||}$ directed along the tube surface.}
    \label{fig:tube}
\end{figure}

 The particle-accelerating component of the electric field, directed along the initial guiding line of the magnetic tube, is related to the plasma velocity component, orthogonal to the lateral surface of the tube, and the azimuthal component of the magnetic field in the following way (see Figure \ref{fig:tube}):
\begin{align}
    \label{electr_field}
    E_{\Vert}=-\frac{1}{c}v_{n}B_{\phi}.
\end{align}
\par From this one can see that in the absence of the velocity components $v_{r}$ and $v_{z}$, which is the case in the linear approximation made in Section \ref{sec2} when analysing the propagation of the Alfvén pulse, the longitudinal component of the electric field is also absent and hence no plasma flow or particle acceleration occurs.
Here, we do not take into account generation of the electric field longitudinal component due to electron inertia, electron pressure gradient, Hall effect, and other effects that can arise in consideration beyond the single-fluid ideal MHD theory (see, e.g., \citealp{stasiewiczSMALLSCALEALFVENIC2000,cramerPhysicsAlfvenWaves2001,tsiklauriMissingPiecesSolar2009}). However, in the case of inertial or kinetic Alfvén waves,  as was shown by \cite{zaitsevParticleAccelerationPlasma2015a}, it cannot provide the necessary acceleration rate of particles.

\par As will be shown below, if the amplitude of the Alfvén wave is sufficiently large, a normal component of the plasma velocity arises, leading to the appearance of a longitudinal non-linear component of the electric field, proportional to $B_{\phi}^{3}$. The similar mechanism of generating super-Dreicer electric field by Alfvén waves for the case of planar waveguide with a table-type plasma density profile and a homogeneous magnetic field was demonstrated  by \cite{tsiklauriMechanismElectricMHD2006} on the basis of   MHD simulations. Below we consider an actual configuration of the axially symmetric magnetic tube with essential longitudinal variation of magnetic field and show that this variation is an important feature for the generating the strong electric field parallel to the unperturbed magnetic field lines.   
\par Let us write the Euler equation that determines the change in plasma velocity:
\begin{align}
    \dfrac{\partial \vec{v}}{\partial t}=-\frac{1}{\rho}\nabla P-\frac{1}{4\pi \rho}\Big(\vec{B}\times(\nabla \times \vec{B})\Big).
\end{align}
Projecting this equation onto the normal to the surface of the magnetic flux tube, we obtain:
\begin{align}
    \label{norm_velocity}
    \dfrac{\partial v_{n}}{\partial t}=-\frac{1}{\rho}\frac{\partial P}{\partial n}-\frac{1}{ \rho}\frac{\partial}{\partial n}\Big(\frac{B^2}{8\pi}\Big)+\frac{B^2}{4\pi\rho}\frac{1}{R}.
\end{align}
Here ${\partial}{/\partial n}=(\hat{\bf{n}} \cdot\nabla)$, and $R$ is the radius of the magnetic field lines curvature. The vector $\hat{\bf{n}}$ is the normal vector to the magnetic tube surface with the following components in the cylindrical coordinate system:
\begin{align}
    \hat{\bf{n}}=\Big(-\frac{B_{r}}{\sqrt{B_{r}^2+B_{z}^2}},0,\frac{B_{z}}{\sqrt{B_{r}^2+B_{z}^2}}\Big).
\end{align}
In the stationary case or at small amplitude of the Alfvén wave, we have
\begin{align}
    -\frac{1}{\rho}\frac{\partial P}{\partial n}-\frac{1}{ \rho}\frac{\partial}{\partial n}\Big(\frac{B^2}{8\pi}\Big)+\frac{B^2}{4\pi\rho}\frac{1}{R}=0.
\end{align}
\par To estimate the strength of the electric field magnitude, we assume that the magnitude of the azimuthal component of the pulse magnetic field, $B_{\phi}$, is sufficiently large so that a change in the magnetic pressure cannot be compensated by a change in the gas pressure, $P \ll {B_{\phi}^2}/{8\pi}$. Suppose also that the characteristic scale of the Alfvén pulse is much smaller than the characteristic curvature radius, $R$, of the unperturbed magnetic field lines. Then, from Equation $\ref{norm_velocity}$ we have an approximate equation for the normal component of the plasma velocity: 
\begin{align}
    \dfrac{\partial v_{n}}{\partial t} \approx -\frac{1}{ \rho}\frac{\partial}{\partial n}\Big(\frac{B_{\phi}^2}{8\pi}\Big).
\end{align}
\par Let us use the solution for the Alfvén wave in the case of an exponentially expanding tube, Equation \ref{expon_sol}, found in the previous section. Note that for the chosen magnetic tube ($\Phi=const$) the axial component of the magnetic field $B_{\phi}$ depends only on time as $B_{\phi}(z-c_{A}t)$ (for a short pulse with a characteristic length $\lambda \lesssim 4H$ the exponential multiplier can be ignored, see Equation \ref{expon_sol}). Then we can replace the differentiation on the time variable $t$ by the differentiation on $z$:
\begin{align}
\label{approx_v_n}
    -c_{A}\dfrac{\partial v_{n}}{\partial z} \approx -\frac{1}{ \rho}\frac{\partial}{\partial n}\Big(\frac{B_{\phi}^2}{8\pi}\Big).
\end{align}
Integrating Equation \ref{approx_v_n} and taking into account that the normal component of the plasma flow velocity is absent before the pulse arrival, i.e. $v_{n}(+\infty)\equiv0$, we get:
\begin{align}
    \label{norm_vel}
    v_{n} \approx- \int^{+\infty}_{z} \frac{1}{c_{A} \rho}\frac{\partial}{\partial n}\Big(\frac{B_{\phi}^2}{8\pi}\Big) dz'.
\end{align}
Substituting this expression into the formula for the electric field longitudinal component, Equation \ref{electr_field}, we finally obtain the following approximate relation:
\begin{align}
    E_{\Vert}\approx\frac{1}{c}B_{\phi}\int^{\infty}_{z} \frac{1}{c_{A} \rho}\frac{\partial}{\partial n}\Big(\frac{B_{\phi}^2}{8\pi}\Big) dz'.
\end{align}
Note the strong non-linear dependence of the velocity component $v_{n}$ and the electric field component $E_{||}$ on the azimuthal component $B_{\phi}$ of the Alfvén-pulse magnetic field.
\par For order of magnitude estimation, let ${\partial}/{\partial n} \approx 1/\delta$, where $\delta$ is the characteristic transverse scale of the Alfvén pulse, and consider that, for a localised pulse with characteristic length $\lambda$, the main part of the integral \ref{norm_vel} is yielded by the interval $(c_{A}t-\lambda/2,c_{A}t+\lambda/2)$. As a result, we can write approximately
\begin{align}
    E_{\Vert}\approx\frac{1}{c}\frac{\lambda}{\delta}\frac{B_{\phi}^3}{8\pi c_{A}\rho}.
\end{align}
Introducing the parameter $\alpha=({B_{\phi}}/{B_{z}})^3$, we rewrite this expression as follows
\begin{align}
    |E_{\Vert}|\approx \Big|\frac{\lambda}{\delta}\frac{c_{A}}{c}\frac{\alpha}{2}B_{z}\Big|.
\end{align}
Again using the solution for the case of an exponentially expanding magnetic tube, Equation \ref{expon_sol}, considering the dependence of the magnetic field components $B_{\phi}$, $B_{z}$ and the characteristic transverse pulse scale $\delta\sim a(z)$ (see Equation \ref{radius}) on the longitudinal $z$-coordinate, and assuming that the electric current distribution is homogeneous over the cross section of the tube, we obtain
\begin{align}
    \label{El_f_val}
   |\langle{E}_{\Vert}\rangle|\approx\Big|\frac{\lambda}{\delta_{0}}\frac{c_{A}}{c}\frac{\langle{\alpha}_{0}\rangle B_{z0}}{10}\Big|.
\end{align}
Here the brackets denote the averaging over the magnetic tube cross-section and the lower index $0$ means the values of the corresponding quantities at the lowest point of the tube, i.e., at $z=0$. As can be seen from the above formula \ref{El_f_val}, the absolute value of the non-linear component of the longitudinal electric field does not follow the exponential decrease in the amplitude of the Alfvén wave $B_{\phi}$ (see Equation \ref{expon_sol}) and could change only slowly, e.g., due to the longitudinal dependence of the Alfvén velocity.
\par Dividing the obtained expression by the value of the Dreicer electric field (\citealp{zaitsevRayleighTaylorInstability2016a}) gives:
\begin{align}
\label{Ed_ratio}
   \frac{|\langle{E}_{\Vert}\rangle|}{E_{D}}\approx 10^7\frac{\lambda}{\delta_{0}}\frac{c_{A}T\langle{\alpha}_{0}\rangle B_{z0}}{6 n_{0}}\exp\Big({\frac{z}{H}}\Big).
\end{align}
Here the plasma temperature, $T$, Alfvén velocity, $c_{A}$, and pulse scale, $\lambda$, are assumed independent on the vertical $z$-coordinate. Also, an exponential dependence of the Dreicer field proportional to electron number density inside the magnetic tube, $n(z)=n_{0}\exp(-z/{H})$, is taken into account.    
\par At the values of electric currents about $I\approx 10^{10}$ A, magnetic field $B_{z0}\approx 10^{2}-10^{3}$ G, height $H\sim$ 300 km, scales $\lambda\sim\delta_{0}\sim$ 30-300 km, electron number density $n_{0}\approx 10^{11}-10^{13}$ cm$^{-3}$ and plasma temperature $T \approx 10^{4}$ K characteristic for the chromospheric part of the loop, the ratio \ref{Ed_ratio} can already exceed unity at $z=0$, which is necessary for the generation of runaway electrons. However, the tube expansion leads to an additional exponential factor $\sim \exp({{z}/{H}})$, which is absent for the magnetic tube of cylindrical geometry with a constant cross-section considered by \cite{zaitsevRayleighTaylorInstability2016a}. This factor at characteristic heights of the medium chromospheric layer $\sim$ 2000 km (\citealp{avrettModelsSolarChromosphere2008}) can increase the ratio \ref{Ed_ratio} by a factor of $10^{2}-10^{3}$. In particular, if the condition ${\langle{E}_{\Vert}\rangle}/{E_{D}} \geqslant 1$ is fulfilled in the lower part of the chromospheric flux tube, it is moreover fulfilled in the higher region. 
\par According to \cite{zaitsevParticleAccelerationPlasma2015a}, for the homogeneous magnetic tube configuration the particles energy of the order of 1 MeV is expected due to acceleration at a distance of about 100 km in a non-linear electric field with a strength slightly higher than the Dreicer field, which is equal to 0.1 V cm$^{-1}$, at a plasma temperature $T \approx 10^{4}$ K and number density $n\approx 10^{11}$ cm$^{-3}$. However, as can be seen from our results for the barometric expanding model of the chromospheric tube, the energy of the accelerated particles can be much higher, reaching values of the order of 1 GeV, since already in the middle chromosphere the electric field can significantly exceed the Dreicer limit. The continuum gamma-ray emission and radiation from the decay of neutral pions recorded during the most powerful solar flares indicate that particles with such energies up to 1 GeV are produced (\citealp{ millerCriticalIssuesUnderstanding1997,benzFlareObservations2017,Ajello_2021}), and their direct acceleration by the electric field can explain the phenomenon under the aforementioned conditions.
\section{Conclusions}
    \label{sec5} 

\par The paper presents an analytical study of the propagation of the Alfvén pulse in an axially symmetric magnetic tube of variable diameter. In the linear approximation for a wide range of parameters, it is found that the problem is reduced to the oscillations of a one-dimensional string and has a self-similar solution. Based on this solution, by means of the perturbation theory, it is shown that an Alfvén pulse of sufficiently large amplitude has a non-linear electric field component, which is directed along the magnetic field of the tube, i.e., along its guiding line, and can inject the chromospheric plasma and/or high energy particles into the coronal regions of the loop. 
\par The obtained result not only confirms the conclusions gained by \cite{zaitsevParticleAccelerationPlasma2015a,zaitsevRayleighTaylorInstability2016a}, but also shows that in an exponentially expanding magnetic tube the excess of this electric field over the Dreicer field, ${\langle{E}_{\Vert}\rangle}/{E_{D}}$, increases in the direction of decreasing plasma density (i.e, from the photosphere to the corona) and can reach values of the order of $10^{2}-10^{3}$. Consequently, due to inhomogeneity of the magnetic field the chromospheric part of coronal loops are favourable to meet the conditions for the generation of the super-Dreicer electric field by Alfvén pulses. This field can accelerate electrons to energies up to the order of 1 GeV and ensure that the coronal loops are filled with a large number of such high-energy electrons. 
\par Note that the general linear solution obtained in Section \ref{sec2} for the propagating Alfvén pulse is applicable for the arbitrary longitudinal profiles of the tube parameters and shear of magnetic field lines. However, the non-linear estimates of the accelerating electric field value provided in    Section \ref{sec4} are valid only for small deviations of the chromospheric tube structure from the exponential law of the barometric model. Consideration of the non-linear phenomena in a twisted flux tube with strong shear of the magnetic field lines owing to a valuable longitudinal current have been carried out during last ten years, presumably starting from the work  of \cite{murawskiNumericalModelMHD2015}, and is still challenging. Such a non-linear consideration is model dependent and requires reasonable analysis of an actual structure of the twisted magnetic tubes with a non-exponential longitudinal profile (cf. a misleading model of the twisted flux tube in a recent work of \citealp{pradhanLinearAnalysisTorsional2024}, where a basic law of the divergence-free magnetic field is violated).  
\par A quantitative solution to the problem of the energetic particle injection into coronal loops requires a detailed analysis of the characteristic parameters and the generation efficiency of Alfvén pulses in the lower chromosphere in quiescent, active and pre-flare regions. Analytical and numerical calculations of the acceleration and escape kinetics of high-energy electrons in the electric field found for an Alfvén pulse of a given amplitude and duration are also required. Finally, it is important to study the non-linear damping of the Alfvén pulse due to its decay under the feedback of runaway electrons. In addition, it is of particular interest to study the reflection and passage of Alfvén pulses in the region of the transition layer between the chromosphere and the corona. Among other things, it is challenging to analyse the generation of strong small-scale kinetic turbulence, appearing due to injection of accelerated electrons by the super-Dreicer electric field, since that turbulence can contribute to a large increase in the anomalous conductivity and effective temperature of plasma, i.e., to heating of the corona as a whole.

\begin{acks}
  The authors are grateful to V.V. Zaitsev for the discussion of the obtained results. This work was supported by a RSF grant No. 22-12-00308.
\end{acks}
\begin{conflict}
The authors declare that they have no conflicts of interest.
\end{conflict}
  

\bibliography{biblio}

\end{document}